\documentclass[preprint,11pt]{elsarticle}
\usepackage[utf8]{inputenc}
\usepackage{amsmath}
\usepackage{amsfonts}
\usepackage{amssymb}
\usepackage{graphicx} 
\usepackage{floatrow}
\usepackage[dvipsnames]{xcolor}
\usepackage{makecell}
\usepackage[version=4]{mhchem}
\usepackage[per-mode=symbol,round-precision=3,scientific-notation=false,range-phrase = \text{--}]{siunitx}
\usepackage[margin=2cm,font=footnotesize]{caption}
\usepackage[left=0.5in,right=0.5in,top=1in,bottom=1in]{geometry}
\pdfoutput=1 

\usepackage{floatrow}
\usepackage{framed} 
\usepackage{multicol} 
\usepackage{nomencl} 
\usepackage{ifthen}
\renewcommand\nomgroup[1]{%
  \ifthenelse{\equal{#1}{A}}{%
    \item[\textbf{Acronyms}]}{
  \ifthenelse{\equal{#1}{R}}{%
    \item[\textbf{Roman Symbols}]}{
  \ifthenelse{\equal{#1}{G}}{%
    \item[\textbf{Greek Symbols}]}{
  \ifthenelse{\equal{#1}{S}}{%
    \item[\textbf{Superscripts}]}{
  \ifthenelse{\equal{#1}{U}}{%
    \item[\textbf{Subscripts}]}{
  \ifthenelse{\equal{#1}{X}}{%
    \item[\textbf{Other Symbols}]}{
  {}}}}}}}}
\renewcommand*{\nompreamble}{\markboth{\nomname}{\nomname}}

\makenomenclature
\renewcommand*\nompreamble{\begin{multicols}{2}}
\renewcommand*\nompostamble{\end{multicols}}

\newcommand\T{\rule{0pt}{2.6ex}}
\newcommand\B{\rule[-1.2ex]{0pt}{0pt}}

\usepackage{xspace}
\newcommand*{\eg}{e.g.,\@\xspace}
\newcommand*{\ie}{i.e.,\@\xspace}
\newcommand*{\etal}{et al.\@\xspace}
\newcommand*{\vs}{vs.\@\xspace}
\newcommand*{\ch}{Ch.\@\xspace}

\makeatletter
\newcommand*{\etc}{%
    \@ifnextchar{.}%
        {etc}%
        {etc.\@\xspace}%
}
\makeatother

\usepackage[colorlinks=true,linkcolor=blue,citecolor=blue,urlcolor=blue]{hyperref}
\usepackage[nameinlink,capitalize]{cleveref}
\DeclareUnicodeCharacter{025B}{\ensuremath{\varepsilon}}

\graphicspath{{figures/}}

\usepackage{setspace}

\usepackage[section]{placeins}

\newcommand{\mcc}[1]{\makecell{#1}}
\newcommand{\mcl}[1]{\makecell[cl]{#1}}

\newcommand{\pkg}[1]{\texttt{#1}}

\usepackage[scaled=0.86]{helvet}
\newcommand*{\helvet}{\fontfamily{phv}\selectfont}
\newcommand{\htxt}[1]{{\helvet{#1}}}

\usepackage[abs]{overpic}
\usepackage{multirow}

\newcommand{\botf}[1]{\includegraphics[width=2cm,trim={0.5cm 1.6cm 0.3cm 1.7cm},clip]{#1}}
\newcommand{\gk}[1]{$g_{#1}\left(\hat{\alpha}_{ijk}\right)$}

\newcommand{\voidplots}[2]{
\begin{overpic}[width=2cm,trim={0.5cm 1.6cm 0.3cm 1.7cm},clip]{Test-#1/#1_Svoid.pdf} \put(1,33.5){\Large \htxt{#2}} \end{overpic} &&
\botf{Test-#1/F-16_Svoid.pdf} &
\botf{Test-#1/F-6_Svoid.pdf} &
\botf{Test-#1/F-1_Svoid.pdf} & &
\botf{Test-#1/F-14_Svoid.pdf} &
\botf{Test-#1/F-7_Svoid.pdf} &
\botf{Test-#1/F-15_Svoid.pdf} &
\botf{Test-#1/F-10_Svoid.pdf} & &

\botf{Test-#1/F-11_Svoid.pdf} &
\botf{Test-#1/F-2_Svoid.pdf} &
\botf{Test-#1/F-13_Svoid.pdf} &
\botf{Test-#1/F-8_Svoid.pdf} &
\botf{Test-#1/F-3_Svoid.pdf} &
\botf{Test-#1/F-5_Svoid.pdf} &
\botf{Test-#1/F-4_Svoid.pdf} &
\botf{Test-#1/F-9_Svoid.pdf} &
\botf{Test-#1/F-12_Svoid.pdf}
}

\creflabelformat{equation}{#2\textup{#1}#3}

\begin{document}

\begin{frontmatter}
\title{Inference of Gas-liquid Flowrate using Neural Networks}
\author[mit]{Akshay J. Dave\corref{cor}}
\ead{akshayjd@mit.edu}
\cortext[cor]{Corresponding author.}
\address[mit]{Nuclear Reactor Laboratory, Massachusetts Institute of Technology, 77 Massachusetts Avenue, Cambridge, MA 02140.}
\author[mic]{Annalisa Manera}
\address[mic]{Nuclear Engineering and Radiological Science, University of Michigan Ann Arbor, 2355 Bonisteel Blvd, Ann Arbor, MI 48109.}
\journal{International Journal of Multiphase Flow}

\begin{abstract}

The metering of gas-liquid flows is difficult due to the non-linear relationship between flow regimes and fluid properties, flow orientation, channel geometry, \etc. 
In fact, a majority of commercial multiphase flow meters have a low accuracy, limited range of operation or require a physical separation of the phases. 
We introduce the inference of gas-liquid flowrates using a neural network model that is trained by wire-mesh sensor (WMS) experimental data. 
The WMS is an experimental tool that records high-resolution high-frequency 3D void fraction distributions in gas-liquid flows.
The experimental database utilized spans over two orders of superficial velocity magnitude and multiple flow regimes for a vertical small-diameter pipe.
Our findings indicate that a single network can provide accurate and precise inference with below a 7.5\% MAP error across all flow regimes. 
The best performing networks have a combination of a 3D-Convolution head, and an LSTM tail.
The finding indicates that the spatiotemporal features observed in gas-liquid flows can be systematically decomposed and used for inferring phase-wise flowrate.
Our method does not involve any complex pre-processing of the void fraction matrices, resulting in an evaluation time that is negligible when contrasted to the input time-span.
The efficiency of the model manifests in a response time two orders of magnitude lower than the current state-of-the-art.
\end{abstract}


\end{frontmatter}


\section{Introduction}

The importance of accurately characterizing multiphase flow transport transcends several industries. 
In the offshore oil-gas extraction industry, oil-gas-water flows are processed and measurement of the relative composition of each is both a fiscal and often legal requirement \cite{Thorn2013}.
In the chemical and pharmaceutical industry, the accurate measurement of multiphase processes in mixing reactors leads to decreased costs and increased product quality \cite{Ricard2005}.
In electric power generating plants, monitoring steam quality through various components affects plant efficiency and mitigates damage of the components \cite[\ch 1]{viswanathan1989damage}.
Additionally, in light-water nuclear power plants, accurately monitoring two-phase transport has a significant impact on the safety margin \cite[\ch 2]{todreas2011nuclear}.

In general, multiphase flow meters (MFM) have low accuracy, a limited range of operation, or require require physical separation of the phases \cite{Oddie2004}.
Early reviews of MFMs \cite{Rajan1993} indicated that the difficulties stemmed from the non-linear relationship between flow regimes and fluid properties, flow orientation, flow direction, channel geometry, and superficial velocities. 
In fact, a majority of early applications of neural networks to multiphase flows were dedicated to classifying flow regimes.
The experimental tools used to measure multiphase flows are grouped into two categories: intrusive and non-intrusive.
Non-intrusive methods include fast X-ray, gamma-ray and neutron tomography -- methods that are generally used for imaging, classifying flow regimes or quantifying volumetric fractions \cite{Chaouki1997, Heindel2011}. 
Wire-mesh sensors (WMS) fall into the former category and measure electrical impedance across a uniformly distributed array, orthogonal to the flow direction.
WMS have significant benefits over conventional electrical capacitance/magnetic/resistance tomography \cite{Ismail2005}.
WMS provide a high spatiotemporal resolution, do not require complex reconstruction algorithms, and can accommodate varying geometries.
The objective of this study is to utilize wire-mesh sensor recordings of air-water flows and explore the capability of various neural networks in inferring the superficial velocity.
The source code, models, and data samples are available at \href{https://github.com/a-jd/bnn}{github.com/a-jd/bnn}.

\subsection{Previous Work}\label{section_previous}

Early work involved the classification of flows into regimes using various experimental instruments and network types.
Cai \etal \cite{Cai1994} used absolute pressure signals in a horizontal pipe to classify flow regimes using a self-organizing map (SOM). Eight features derived from a \SI{40}{\hertz} pressure signal was used as the input vector. 
Tsoukalas \etal \cite{Tsoukalas1997} used indirect area-averaged impedance-based measurements to identify flow regimes. Flow regime identification required knowledge of superficial velocities, which was deemed difficult at the time.  The authors identified flow regimes manually through flow geometry and trained with a feed-forward neural network (FFN). 
Mi \etal \cite{Mi1998} extended this work to include SOMs to identify flow regimes. 
Mi \etal \cite{Mi2001} further extended this work to train FFNs and SOMs on mixed data (synthetically generated and experimentally obtained impedance measurements) for regime classification.  
Wu \etal \cite{Wu2001} used piezo-differential pressure measurements to classify flow regimes using a multilayer Artificial Neural Network (ANN). The study achieved a greater than 90\% accuracy in the classification of stratified, intermittent and annular flow regimes. 
Hern\'andez \etal \cite{Hernandez2006} and Juli\'a \etal \cite{Julia2008} used a conductivity probe to measure the chord length distribution of each bubble and classify the measurement into a flow regime using FFN/ANN, and SOM, respectively.
Wang \etal \cite{Wang2017} used ANNs to correct the total massflow rate and predict average void fraction using a venturi meter and differential pressure sensor. Wang found that Support Vector Machines achieved a greater reduction of error than ANNs.

A few recent studies have focused on using neural networks to infer the total mass flow rate or superficial velocities in two-phase flows.
Meribout \etal \cite{Meribout2010} used multiple experimental instruments to determine the total mass flow rates of water-oil flows using FFNs and ANNs.
Shaban and Tavoularis used differential pressure signals to infer the superficial gas and liquid velocities using an ANN \cite{Shaban2014}. 
Shaban and Tavoularis used the same experimental facility, but instead using WMS, to achieve the same objective using an ANN \cite{Shaban2015}. 
In both their studies, Shaban and Tavoularis classify the signals by first post-processing the measured time history into few ($\approx10$) features and then use two independent ANNs for predicting the superficial gas and liquid velocities separately ($v_g,v_f$, respectively). 
The post-processing involves statistical binning of the time-history into relative occurrence of differential pressure/area-averaged void fraction, power spectral density of the signal, principal component analysis and finally independent component analysis before connection to the ANNs. 
Furthermore, a separate set of ANNs is trained for each flow regime. 
For four distinct flow regimes, this results in a total of eight independent ANNs. \cref{table_summ} provides a contrast of the literature surveyed.

\begin{table}[!ht]
\caption{Summary of previous studies that use neural networks to address various objectives for multiphase flows. Geometry initialisms: Horizontal Annulus (HA), Vertical Annulus (VA). Instrument initialisms: Area-averaged Impedance (AAI), Conductivity Probe (CP), Differential Pressure (DP), Venturi Meter (VM), Wire-mesh Sensor (WMS). Network initialisms: Artificial Neural Network (ANN), Convolutional Neural Network (CNN), Feed-forward Network (FFN), Self-organizing Map (SOM), Recurrent Neural Network (RNN). }\label{table_summ}
\resizebox{\linewidth}{!}{
\begin{tabular}{@{}lccccc@{}}
\hline\T\B
Author & Objective & Flow & Geometry & Instrument & Network \\
\hline\T\B
\mcl{Cai \etal \\ 1994 \cite{Cai1994}} & Regime Classification & \mcc{$0.192\leq v_g\leq \SI{10.38}{\meter\per\second}$\\$0.71\leq v_f\leq \SI{3.33}{\meter\per\second}$} & \mcc{HA\\$D=\SI{50.0}{\milli\meter}$} & \mcc{Static Pressure\\$f=\SI{40}{\hertz},~T=\SI{102.4}{\second}$} & SOM \\
\hline\T\B
\mcl{Tsoukalas \etal \\ 1997 \cite{Tsoukalas1997}} & Regime Classification & \mcc{$? < v_g < \SI{10}{\meter\per\second}$\\$? < v_f < \SI{10}{\meter\per\second}$} & \mcc{VA\\$D=~?$} & \mcc{AAI\\$f=\SI{100}{\kilo\hertz},~T=~?$} & FFN \\
\hline\T\B
\mcl{Mi \etal \\ 1998 \cite{Mi1998}} & Regime Classification & \mcc{$? < v_g < \SI{50}{\meter\per\second}$\\$? < v_f < \SI{10}{\meter\per\second}$} & \mcc{VA\\$D=\SI{50.8}{\milli\meter}$} & \mcc{AAI\\$f=\SI{100}{\kilo\hertz},~T=~?$} & SOM \\
\hline\T\B
\mcl{Wu \etal \\ 2001 \cite{Wu2001}} & Regime Classification & \mcc{$1.86 < v_g < \SI{29.45}{\meter\per\second}$\\$0.02 < v_f < \SI{0.31}{\meter\per\second}$} & \mcc{HA\\$D=\SI{40.0}{\milli\meter}$} & \mcc{DP\\$f=~?,~T=~?$} & ANN \\
\hline\T\B
\mcl{Hern\'andez \etal \\ 2006 \cite{Hernandez2006}} & Regime Classification & \mcc{$0.0043 \leq v_g\leq \SI{9.83}{\meter\per\second}$\\$0.0314 \leq v_f\leq \SI{2.51}{\meter\per\second}$} & \mcc{VA \\$D=\SI{50.8}{\milli\meter}$} & \mcc{Single CP\\$f=\SI{12}{\kilo\hertz},~T=\SI{60}{\second}$} & FFN, ANN  \\
\hline\T\B
\mcl{Juli\'a \etal \\ 2008 \cite{Julia2008}} & Regime Classification & \mcc{$0.01 \leq v_g\leq \SI{10}{\meter\per\second}$\\$0.03 \leq v_f\leq \SI{2.5}{\meter\per\second}$} & \mcc{VA \\$D=\SI{50.8}{\milli\meter}$} & \mcc{Single CP\\$f=\SI{12}{\kilo\hertz},~T=\SI{60}{\second}$} & SOM  \\
\hline\T\B
\mcl{Meribout \etal \\ 2010 \cite{Meribout2010}} & \mcc{Total Mass Flow\\Rate Inference} & $0\leq\dot{m}\leq\SI{800}{\liter\per\minute}$ & \mcc{VA\\$D=\SI{50.8}{\milli\meter}$} & \mcc{DP, VM, \etc \\ $f=\SI{500}{\hertz},~T=~?$} & FFN, ANN  \\
\hline\T\B
\mcl{Wang \etal \\ 2017 \cite{Wang2017}} & \mcc{Total Mass \& Void\\Fraction Inference} & \mcc{$700\leq\dot{m}\leq\SI{14,500}{\kilo\gram\per\hour}$\\$0\leq\alpha\leq0.3$} & \mcc{HA \& VA\\$D=\SI{25.4}{\milli\meter}$} & \mcc{DP \& VM\\$f=30,~50\SI{}{\hertz},~T=\SI{100}{\second}$} & ANN  \\
\hline\T\B
\mcl{Shaban and \\ Tavoularis 2014 \cite{Shaban2014}} & \mcc{Superficial Velocity\\Inference} & \mcc{$0.014 \leq v_g\leq \SI{22}{\meter\per\second}$\\$0.04 \leq v_f\leq \SI{0.4}{\meter\per\second}$} & \mcc{VA\\$D=\SI{32.5}{\milli\meter}$} & \mcc{DP \\$f=\SI{200}{\hertz},~T=\SI{60}{\second}$} & ANN  \\
\hline\T\B
\mcl{Fan and Yan \\2014 \cite{Fan2014}} & \mcc{Superficial Velocity\\Inference} & \mcc{$0.58 \leq v_g\leq \SI{1.86}{\meter\per\second}$\\$0.35 \leq v_f\leq \SI{1.62}{\meter\per\second}$} & \mcc{HA\\$D=\SI{50.0}{\milli\meter}$} & \mcc{Dual CP\\$f=7,~13\SI{}{\kilo\hertz},~T=\SI{10}{\second}$} & ANN  \\
\hline\T\B
\mcl{Shaban and \\ Tavoularis 2015 \cite{Shaban2015}} & \mcc{Superficial Velocity\\Inference} & \mcc{$0.04 \leq v_g\leq \SI{20}{\meter\per\second}$\\$0.13 \leq v_f\leq \SI{3.0}{\meter\per\second}$} & \mcc{VA\\$D= \SI{32.5}{\milli\meter}$} & \mcc{Dual WMS \\ $f=\SI{1}{\kilo\hertz},~T=\SI{75}{\second}$} & ANN  \\
\hline\T\B
\mcl{This work} & \mcc{Superficial Velocity\\Inference} & \mcc{$0.0233 \leq v_g\leq \SI{4.97}{\meter\per\second}$\\$0.0102 \leq v_f\leq \SI{2.55}{\meter\per\second}$} & \mcc{VA\\$D=\SI{50.8}{\milli\meter}$} & \mcc{Single WMS \\ $f=\SI{2.5}{\kilo\hertz},~T=\SI{0.2}{\second}$} & \mcc{ANN, CNN,\\ RNN}  \\
\hline
\end{tabular}
}
\end{table}

\subsection{Objectives}

The main objective of this work is to investigate the performance of state-of-the-art neural network architectures for inference of superficial velocities in air-water flows using WMS.
In particular, this work will only utilize a single network to output ($v_g,v_f$) as a vector, for all flow regimes. 
Using a single network improves deployability as a multiphase flow-meter.
The WMS signal will be directly analyzed by the network.
In this way any post-processing (requires subjective user intervention) is eliminated, and the architecture of the network has a greater impact on the accuracy of the model.
A salient outcome of this study will be to demonstrate that Convolutional Neural Networks (CNN) and Recurrent Neural Networks (RNN) can provide an improvement in regression of complex non-linear spatiotemporal features encountered in multiphase flows.

\section{Data \& Network Architecture}

\subsection{Experimental Data}

\begin{figure}[!ht] \centering
\resizebox{\linewidth}{!}{
\begin{tabular}{ccccccc}
\begin{overpic}[height=7.5cm]{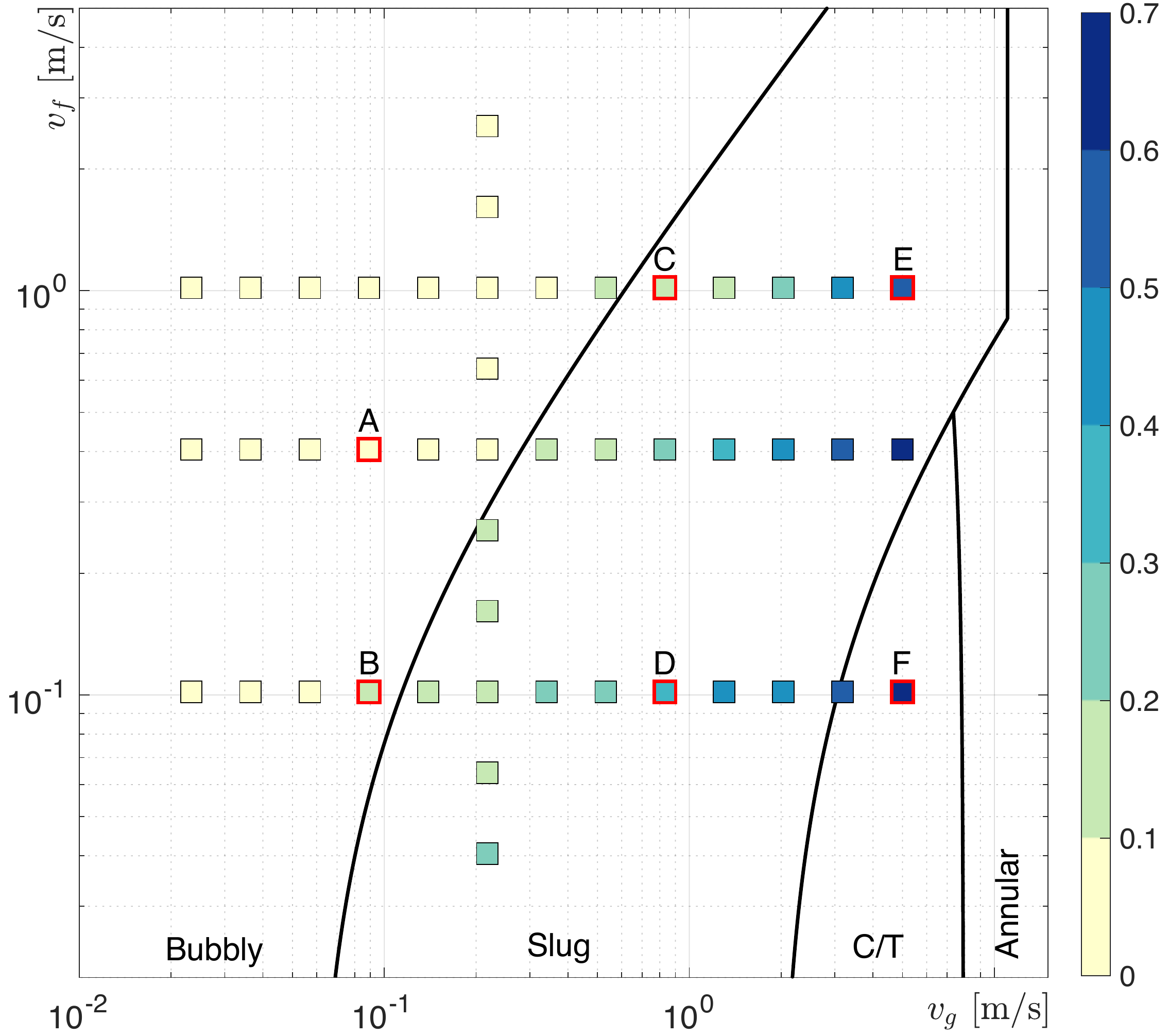} \put(27,198){$\alpha\left[-\right]$} \end{overpic} &
\begin{overpic}[height=7.5cm]{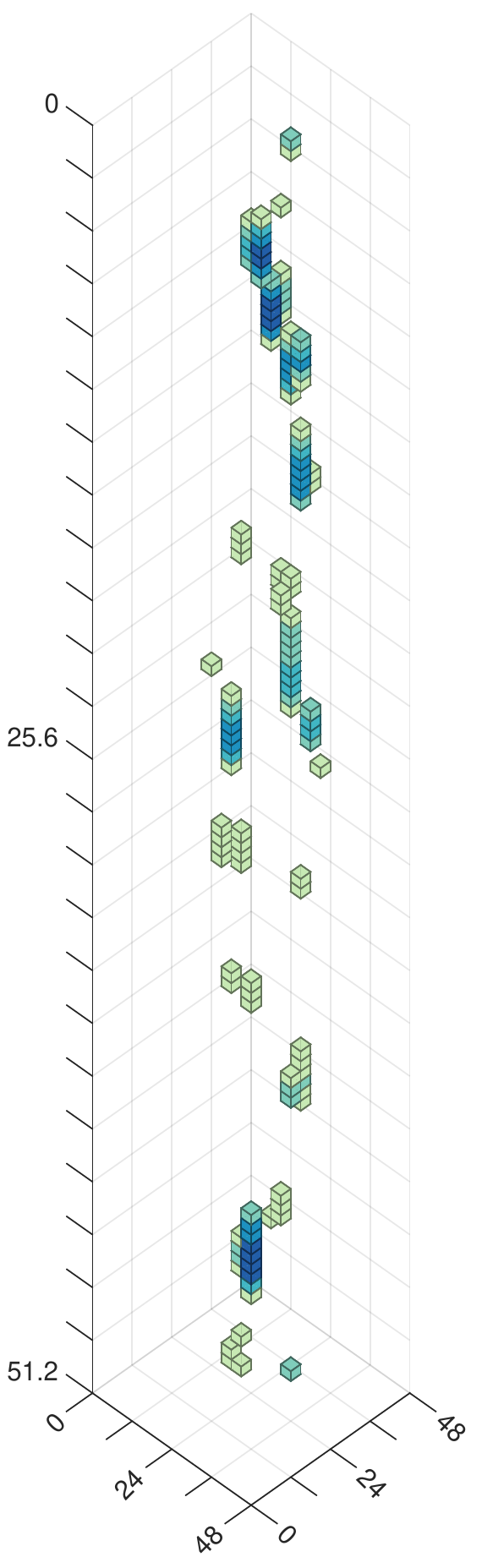} \put(50,205){\htxt{A}} \end{overpic} &
\begin{overpic}[height=7.5cm]{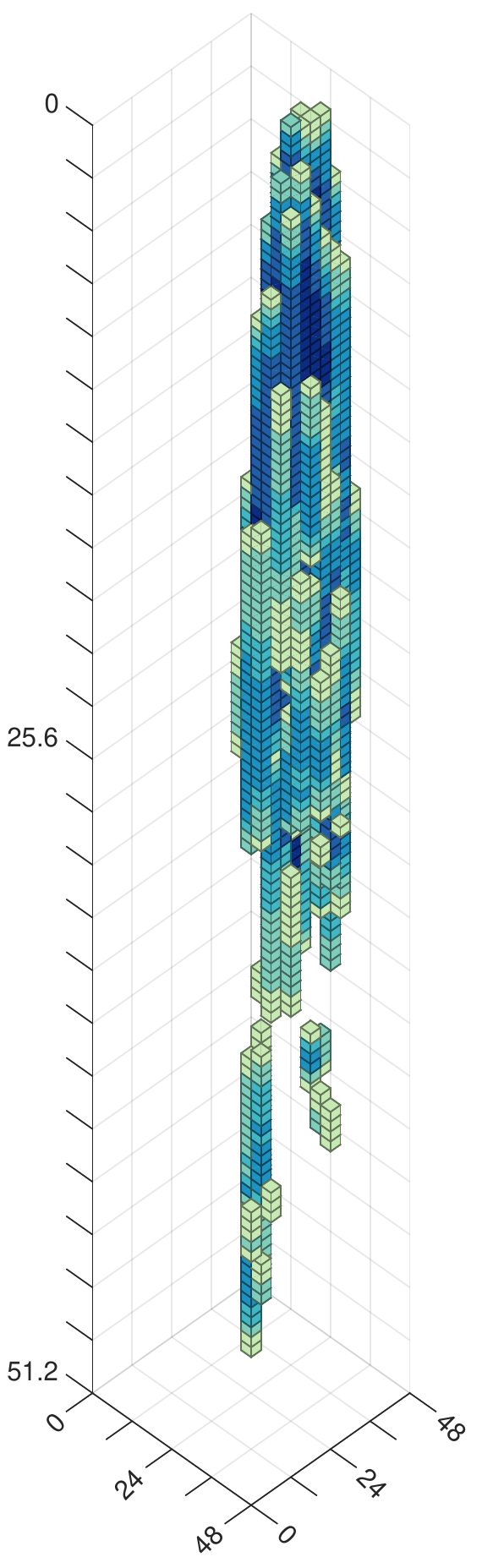} \put(50,205){\htxt{B}} \end{overpic} &
\begin{overpic}[height=7.5cm]{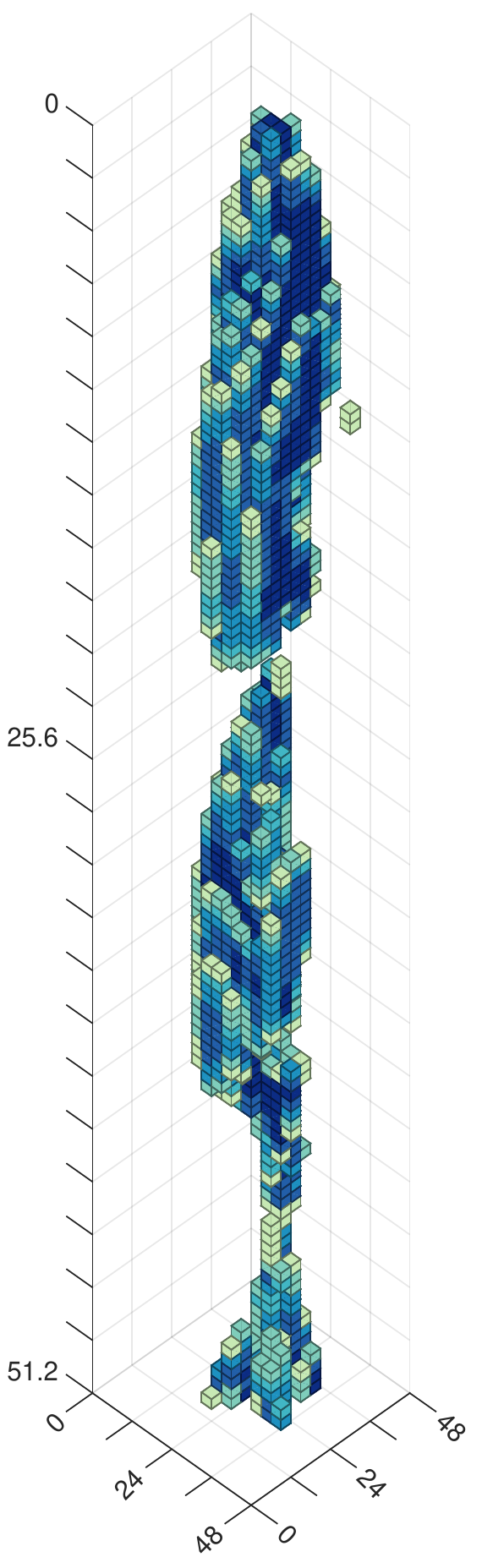} \put(50,205){\htxt{C}} \end{overpic} &
\begin{overpic}[height=7.5cm]{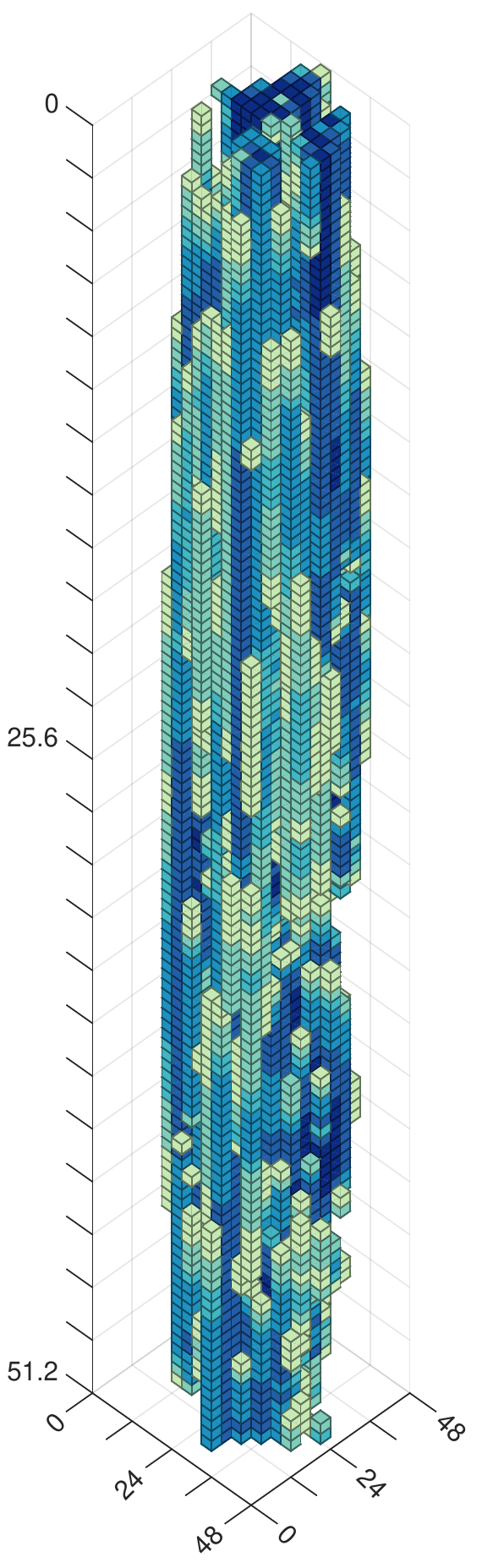} \put(50,205){\htxt{D}} \end{overpic} &
\begin{overpic}[height=7.5cm]{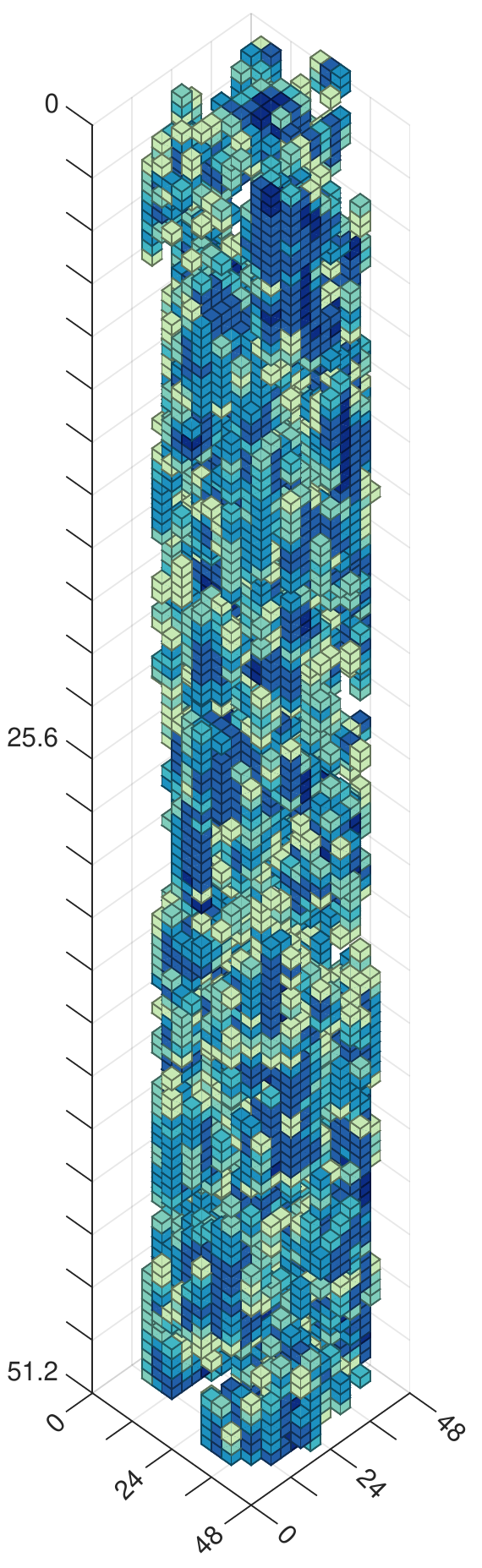} \put(50,205){\htxt{E}} \end{overpic} &
\begin{overpic}[height=7.5cm]{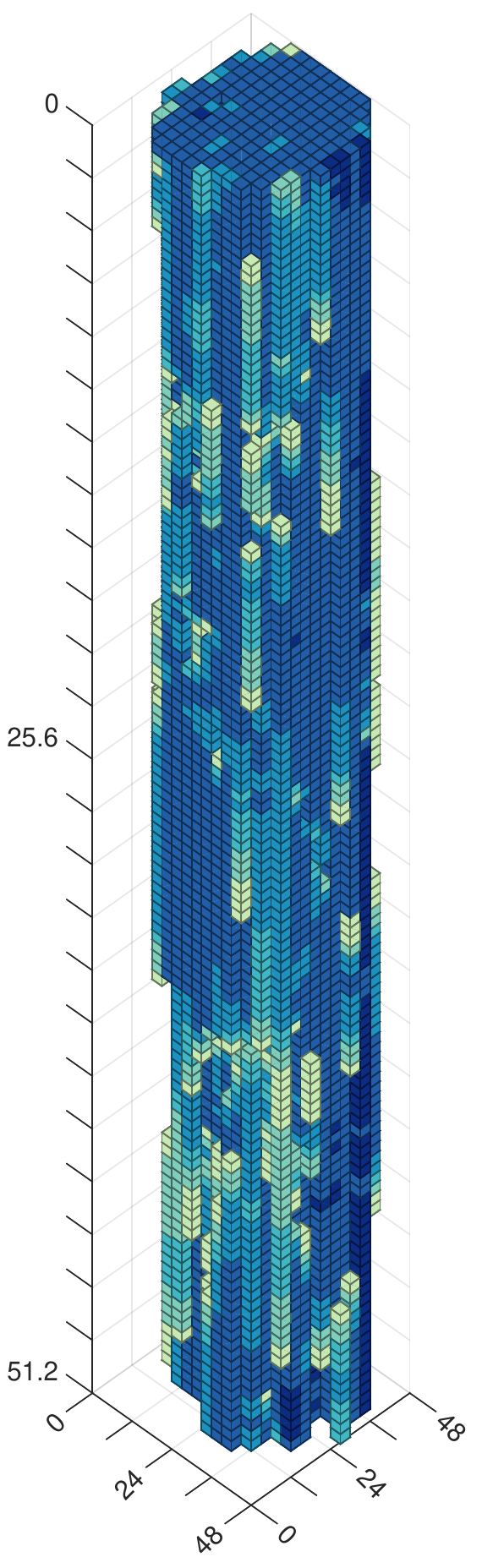} \put(50,205){\htxt{F}} \end{overpic}
\end{tabular}
}
\caption{An overview of the experimental wire-mesh sensor database used in this work. A total of 46 permutations of $(v_g,v_f)$ are included in the training dataset. The left-hand side graph indicates the volume-averaged void fraction ($\alpha$) of each test-condition. The continuous lines indicate \textit{expected} flow regimes \cite{Kaichiro1984}. Voxel plots of void fraction distributions for selected tests are presented on the right-hand side. The $x$-$y$ axes are in millimeters and the $z$ axis is in milliseconds (a total of 128 frames are presented, or \SI{51.2}{\milli\second}).}\label{fig_expdata}
\end{figure}

The experimental wire-mesh sensor (WMS) \cite{prasser1998new} data was obtained from the TOPFLOW facility at Helmholtz-Zentrum Dresden-Rossendorf \cite{prasser2007construction}. 
The experimental facility consists of a vertical \SI{8.0}{\meter} long ($L$), \SI{50.8}{\milli\meter} inner diameter ($D$) annular test section. 
Flow is in the upwards direction.
Multiple flanges allow placement of WMS assembly at 1.9, 31, 59 or 151 $L/D$. Inlet pressure is kept constant at \SI{0.25}{\mega\pascal}.
Although TOPFLOW experiments were recorded using dual WMS, this work only utilizes measurements from a single upstream WMS at 151 $L/D$ (to ensure fully-developed flow).
The WMS is two planar arrays of electrodes, separated by \SI{1.5}{\milli\meter}, and placed orthogonal to each other.
Each array consists of 16 electrodes, uniformly distributed with a spacing of \SI{3.0}{\milli\meter}.
The impedance of electrical signals across the electrodes is calibrated to provide measurements of local void fraction around electrode junctions.
The void fraction is defined as the volumetric ratio of gas to total.
The WMS operates at \SI{2.5}{\kilo\hertz} and records for a total of \SI{10}{\second}.
Therefore, each measurement consists of a $(25000, 16, 16)$ void fraction matrix. 
A summary of the test conditions and representative visualizations of void fraction distributions are presented in \cref{fig_expdata}.

\setlength{\tabcolsep}{0pt}
\setlength{\tabcolsep}{6pt}

Varying liquid and gas flow rates results in varying flow regimes.
As noted in \cref{table_summ}, a plurality of early works were focused on using neural networks to classify flow regimes.
For a \textit{small} diameter pipe \cite{prasser2005influence}, the flows are classified as Bubbly, Slug, Churn-turbulent, and Annular \cite{Kaichiro1984}.
As visualized in \cref{fig_expdata}, each regime has distinct spatiotemporal features.
Therefore, the void fraction matrix, $\alpha_{ijk}$\footnote{The index $i$ represents the temporal dimension. The indices $j,k$ represent the spatial dimensions in the plane of the WMS.}, is not a straightforward function of $(v_g, v_f)$. 
Furthermore, even if we volume-average $\alpha_{ijk}\rightarrow\alpha$, there are multiple sets of $(v_g, v_f)$ observed for each $\alpha$.
The primary objective is to test the capability of a neural network to provide a regression function $\mathbb{F}(\alpha_{ijk})=(v_g, v_f)$.

\subsection{Network Architectures}

\begin{figure}[!ht]
\refstepcounter{figure}\label{fig:arch}
\begin{tabular}{cccccc}
{\helvet Model A} & {\helvet Model B} & {\helvet Model C} & {\helvet Model D} & {\helvet Model E} & {\helvet Model F}\\
\includegraphics[scale=0.45]{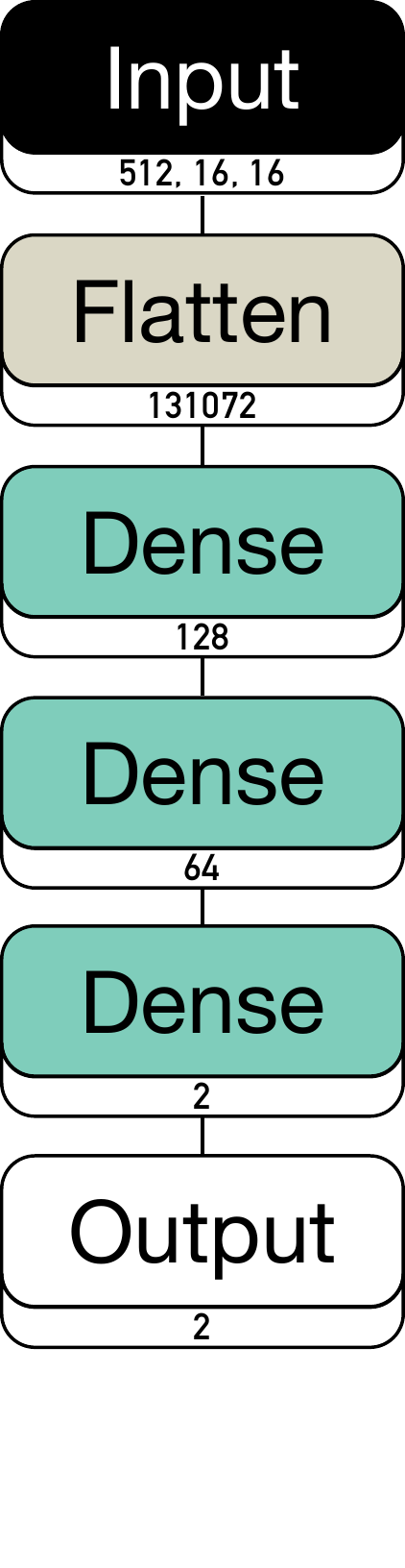} &
\includegraphics[scale=0.45]{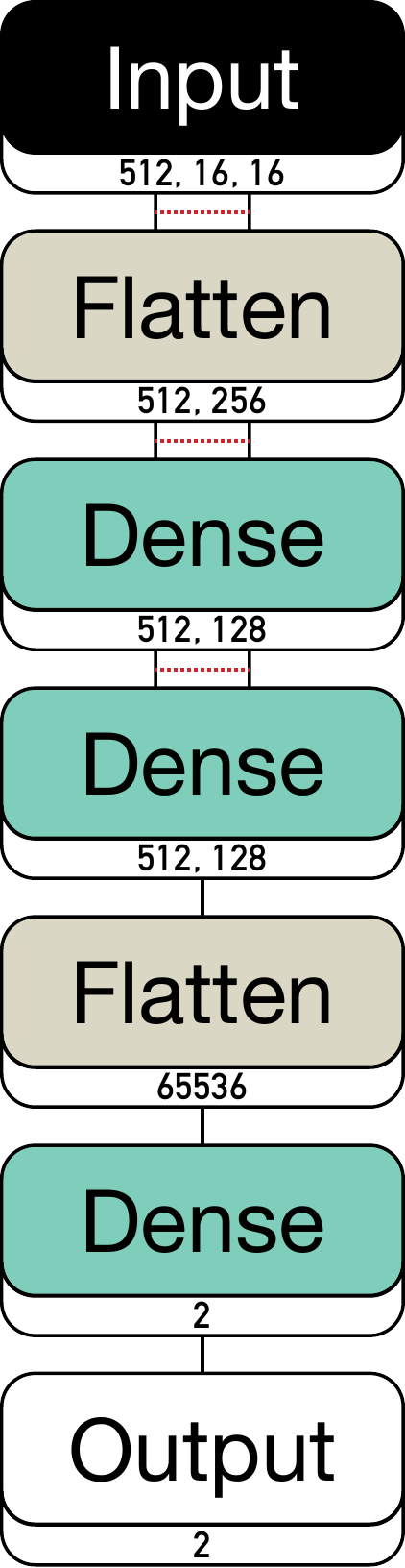} &
\includegraphics[scale=0.45]{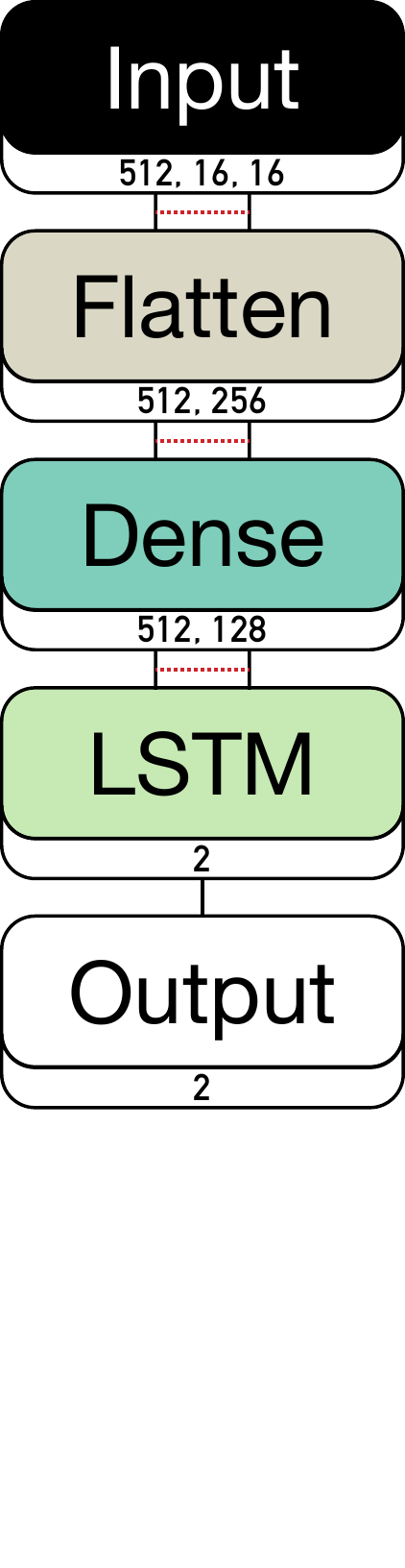} &
\includegraphics[scale=0.45]{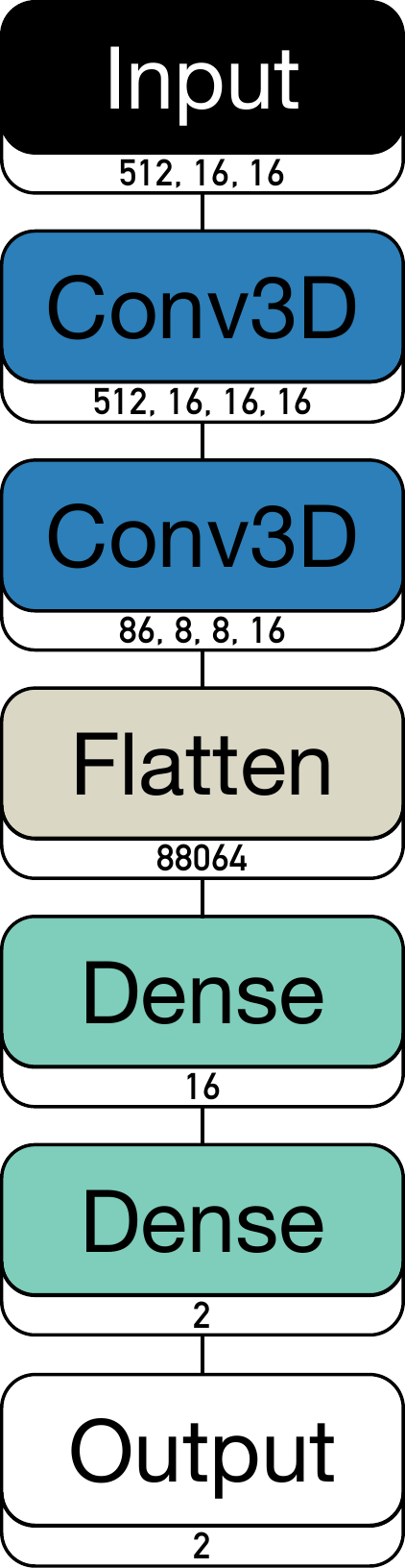} &
\includegraphics[scale=0.45]{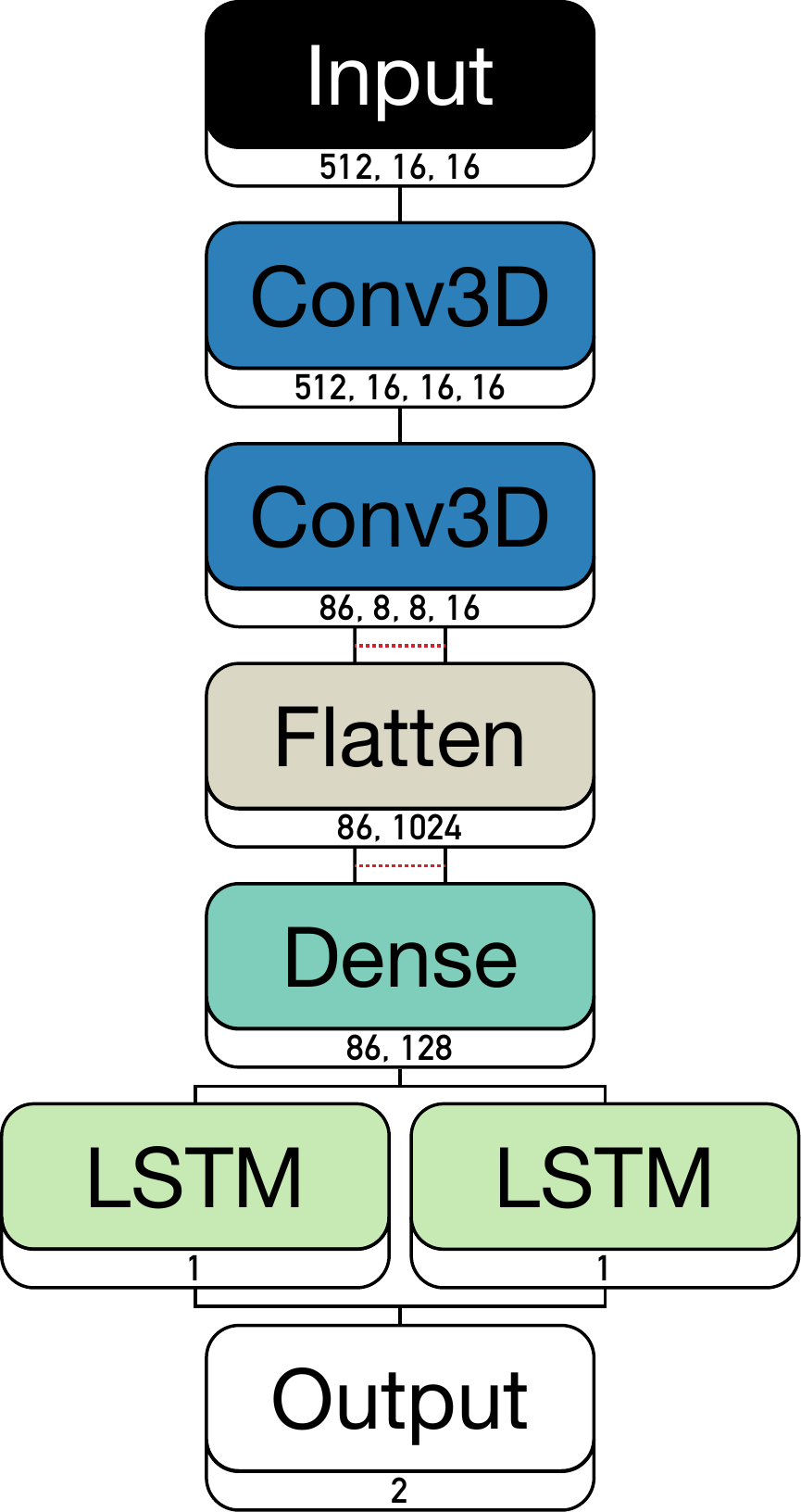} &
\includegraphics[scale=0.45]{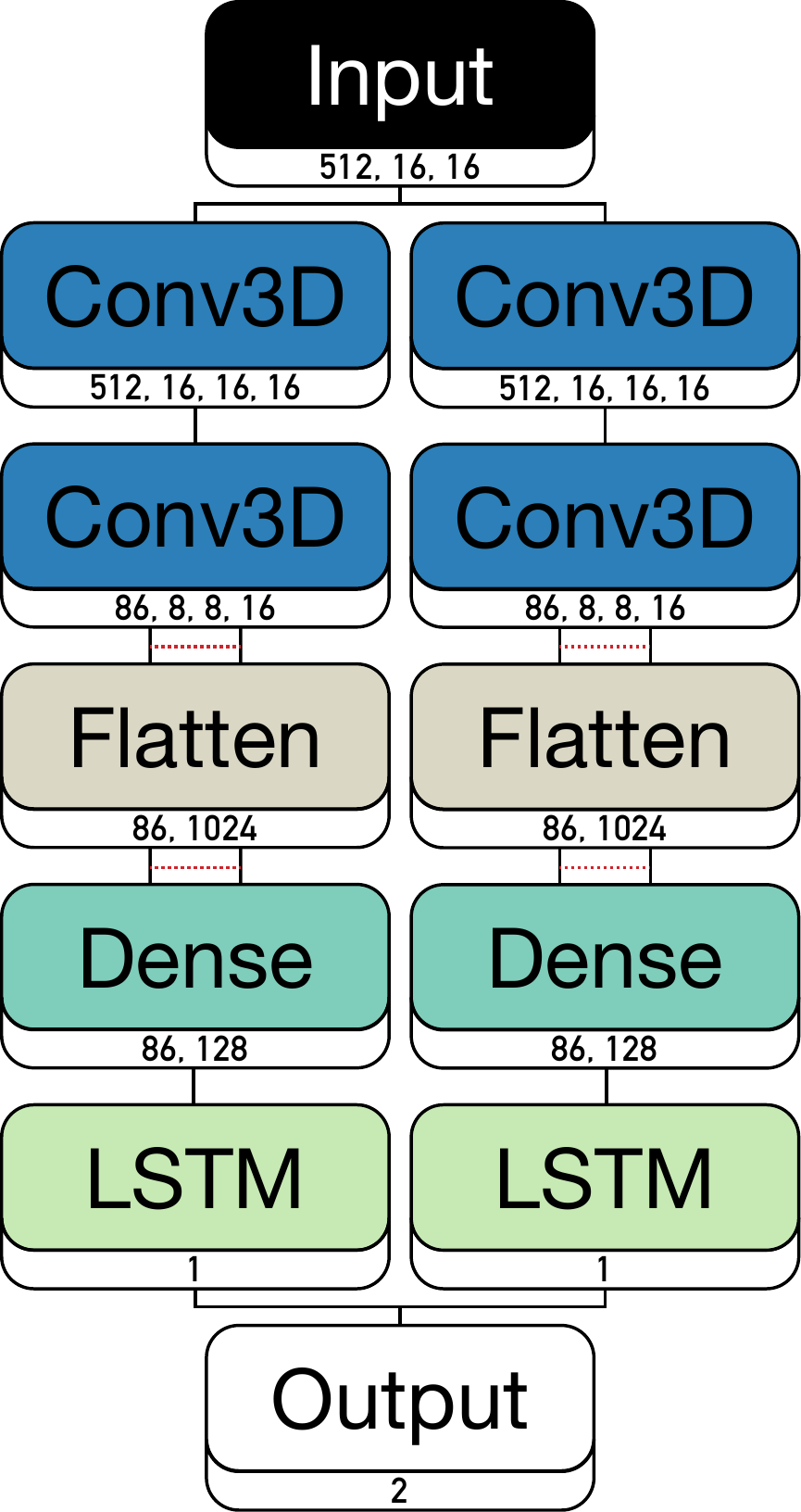} \B \\
\multicolumn{6}{c}{
\begin{tabular}{@{}p{5.5cm}p{4.5cm}@{}}
\footnotesize Figure \ref*{fig:arch}: Neural network architectures. Intermediate output sizes vary based on settings. &
\raisebox{-.7cm}{\fbox{\includegraphics[scale=0.45]{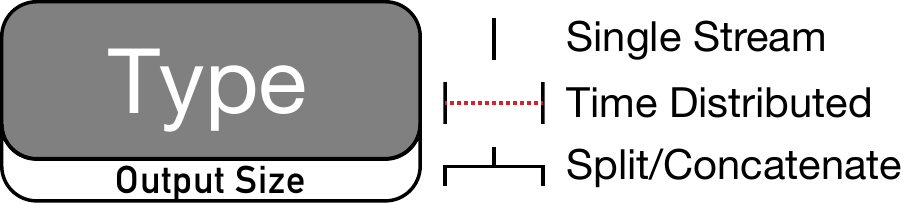}}}
\end{tabular}
}
\end{tabular}

\end{figure}

Deep neural networks consist of several layers of manipulation of the input data.
A summary of the theoretical foundation and best-practices of these layers is found in \cite{goodfellow2016deep}.
This study explored different architectures using the \pkg{Keras} 2.2.4 library \cite{chollet2015keras} (with \pkg{TensorFlow} 1.13 \cite{tensorflow2015} backend). 
The architectures are presented in \cref{fig:arch}. 
They can be separated into two categories: without convolutional heads ({\helvet Model A, B, C}) and with ({\helvet Model D, E, F}).
During the study, various combinations of activation functions, layer size, and convolution parameters were tested.
Using the rectifier \cite{Hahnioser2000} for all intermediate layers resulted in the most successful models.
Since this is a regression problem, the last functional layer did not have an activation function. 

\subsubsection{Data pre-processing}

To facilitate training and adhere to best-practices, both $\alpha_{ijk}$ and $(v_g, v_f)$ were transformed.
The interval of $\alpha_{ijk}$ is $[0, 100]$, as unsigned 8-bit integers.
The interval of $v_g$ is $[0.0234, 4.975]~\SI{}{\meter\per\second}$.
The interval of $v_f$ is $[0.0405, 2.554]~\SI{}{\meter\per\second}$.
Superficial velocities are 32-bit float.
The $\alpha_{ijk}$ matrices are scaled linearly to $[0,1]$.
The $v_g$ and $v_f$ arrays are scaled log-linearly to $[0.1, 0.9]$. The log-linear transformation improved mean average percentage error by 10-20\%. 

Apart from the transformations, the sampling of $\alpha_{ijk}$ is the other major pre-processing step.
There are several reasons $\alpha_{ijk}$ should be sampled from its original size of $(25000,16,16)$.
The first is that there would not be sufficient samples and it would preclude a separate training and validation phase for each $(v_g, v_f)$ permutation.
The second is that to infer a superficial velocity, a large number of frames would be required and therefore the response time of the flow meter would be slow.
Lastly, depending on the architecture chosen, the size of the neural network would quickly balloon.
The size of the sampled void fraction matrix, $\hat{\alpha}_{ijk}$, was $(32\cdot16,16,16)$, corresponding to a measurement time period of \SI{204.8}{\milli\second}.

\section{Results \& Discussion}

\begin{figure}[!ht]
\begin{tabular}{c@{}c}
\begin{tabular}{c@{}c}
\begin{overpic}[height=7.5cm]{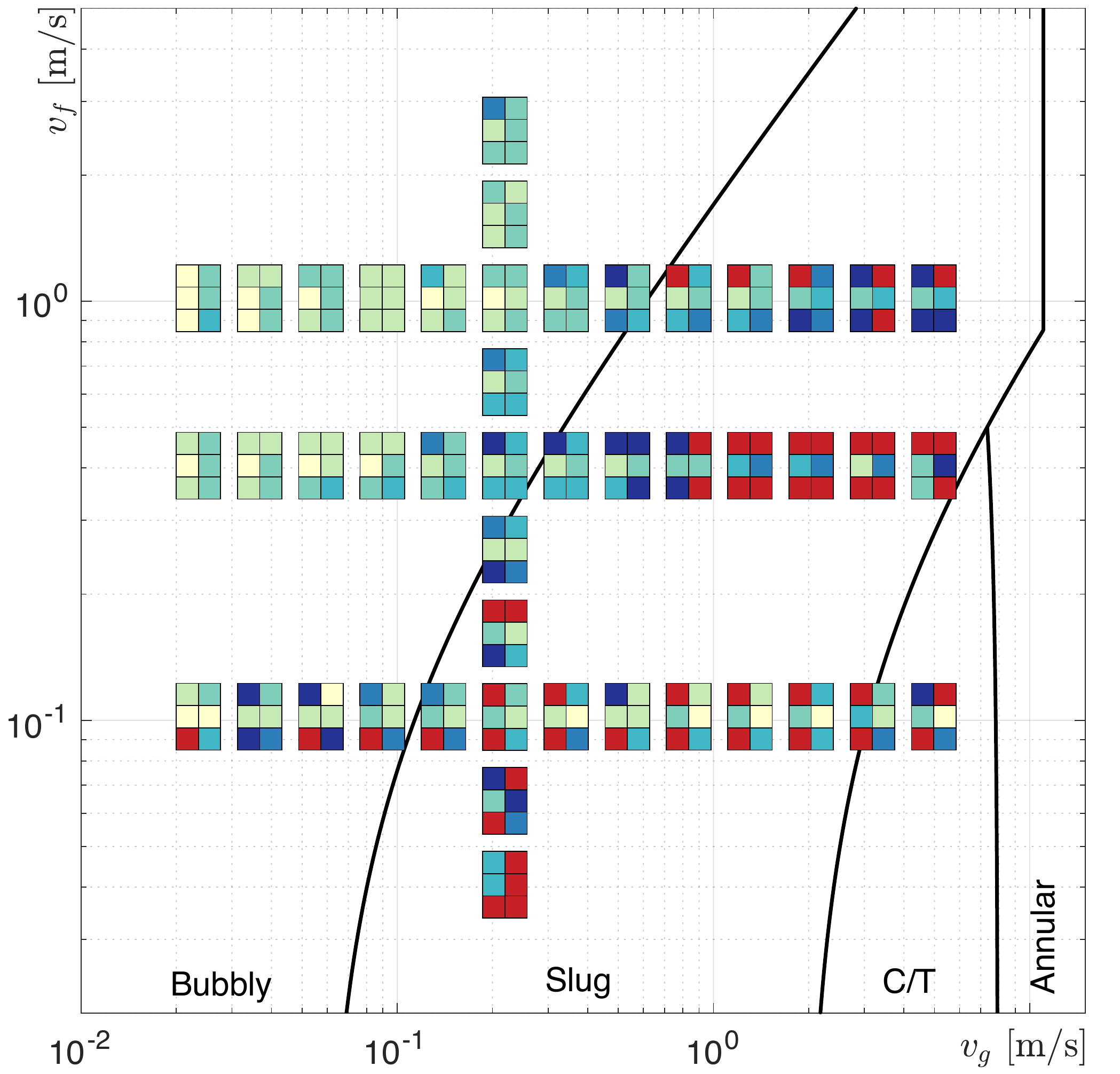} \put(27,198){$\varepsilon\left[\%\right]$} \put(170,170){\setlength{\fboxsep}{0pt}\fbox{\includegraphics[scale=0.35]{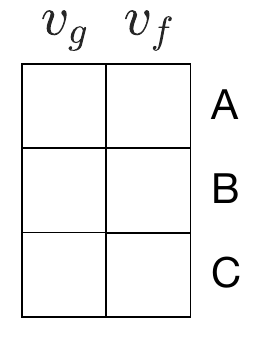}}}\end{overpic} &
\begin{overpic}[height=7.5cm]{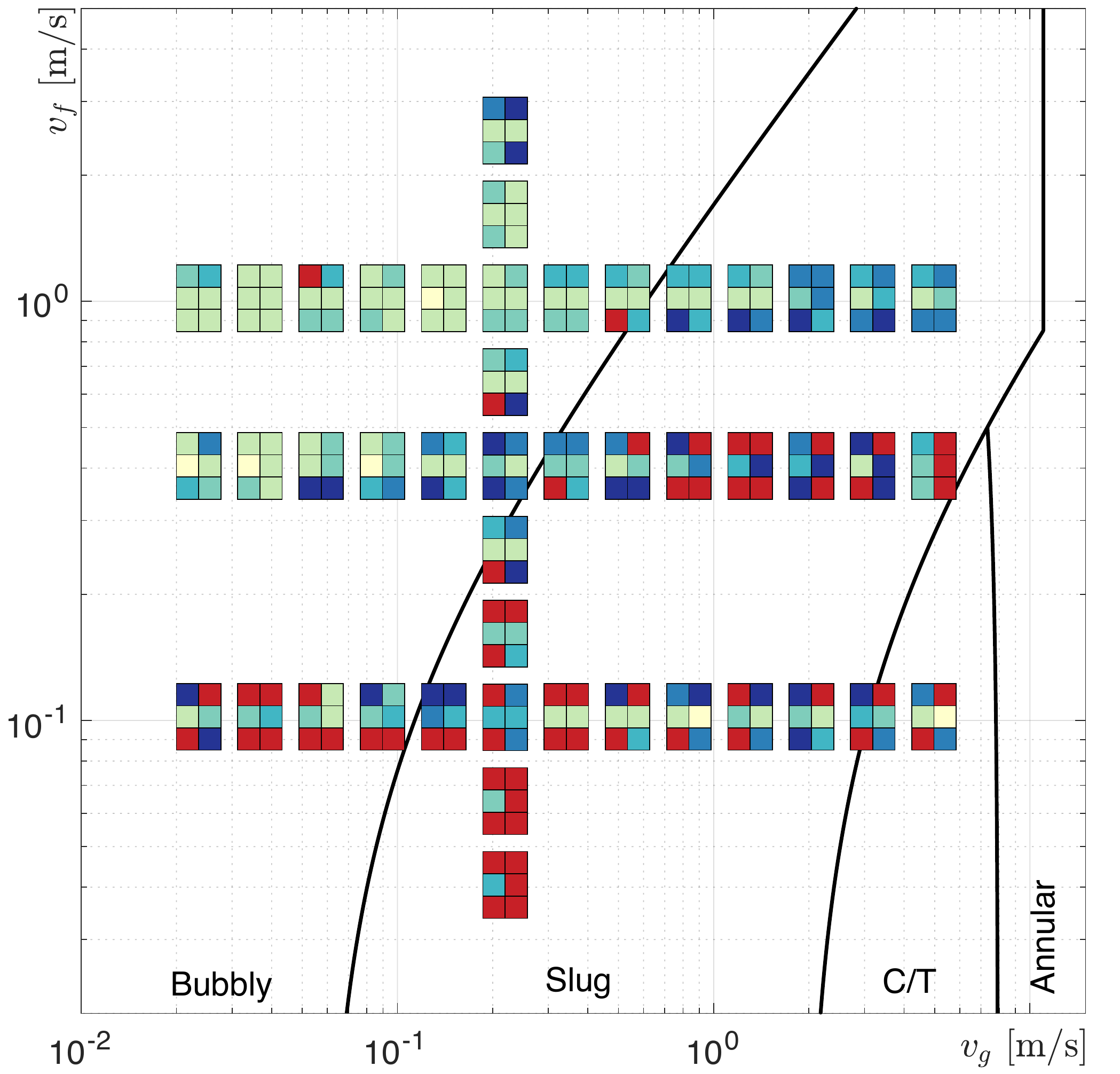} \put(27,198){$\sigma\left[\%\right]$} \put(170,170){\setlength{\fboxsep}{0pt}\fbox{\includegraphics[scale=0.35]{abc-Key.pdf}}}\end{overpic} \B \\
\begin{overpic}[height=7.5cm]{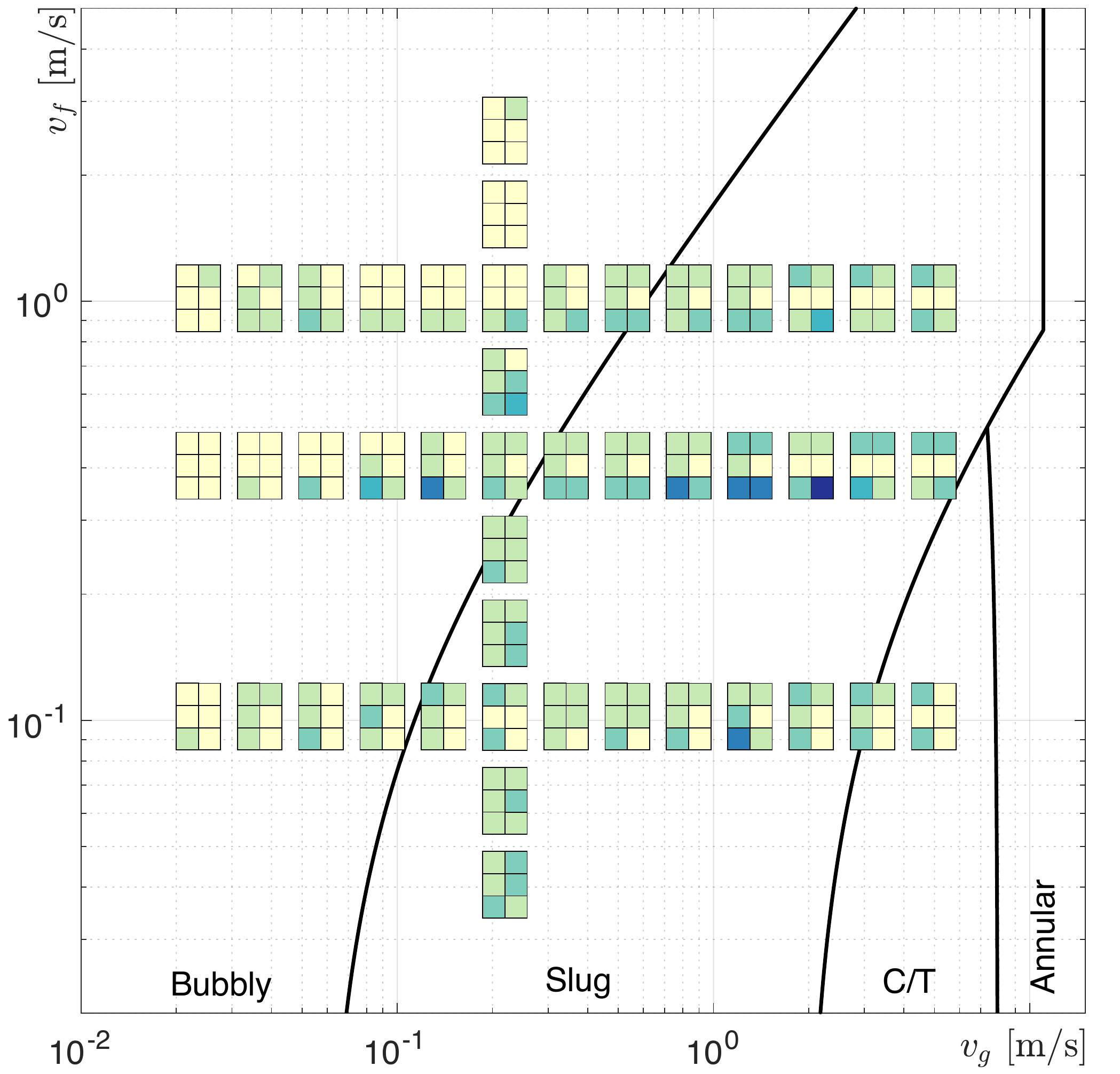} \put(27,198){$\varepsilon\left[\%\right]$} \put(170,170){\setlength{\fboxsep}{0pt}\fbox{\includegraphics[scale=0.35]{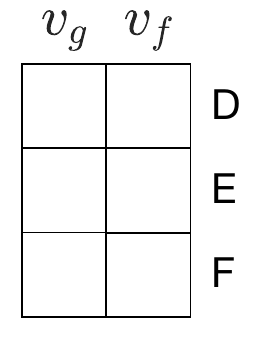}}}\end{overpic} &
\begin{overpic}[height=7.5cm]{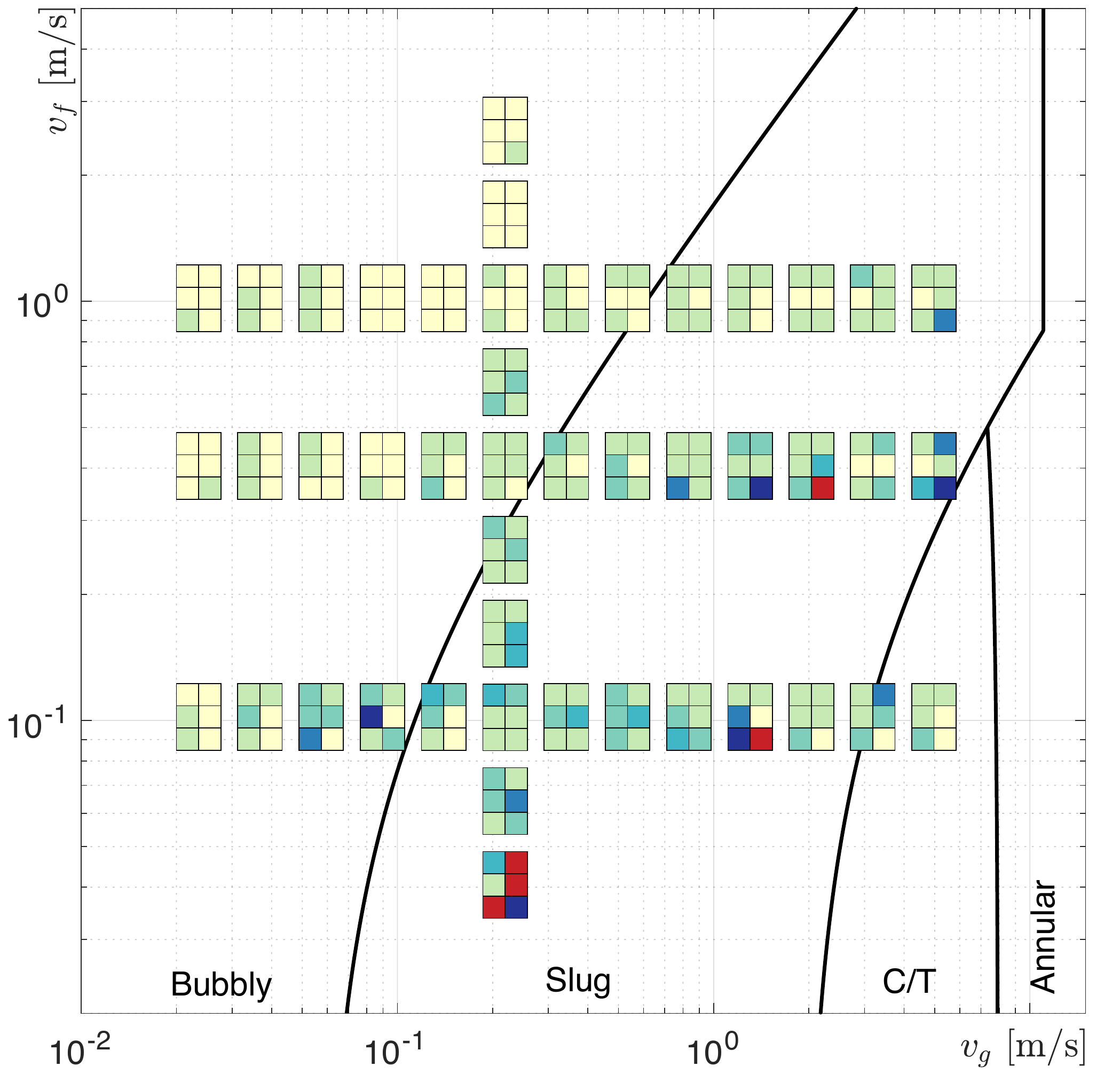} \put(27,198){$\sigma\left[\%\right]$} \put(170,170){\setlength{\fboxsep}{0pt}\fbox{\includegraphics[scale=0.35]{def-Key.pdf}}}\end{overpic} \\
\end{tabular}
& \raisebox{-103pt}{\includegraphics[scale=0.35]{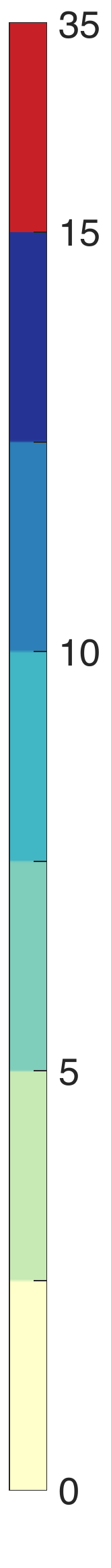}}
\end{tabular}
\caption{Mean average percentage error (left) and corresponding standard deviation (right) for \htxt{Model A-C} (top) and \htxt{Model D-F} (bottom). For each marker, columns represent superficial velocity and rows represent model evaluated. The range of color bars is equivalent for all graphs.}\label{fig:results}
\end{figure}

During the study, several metrics were explored to drive the training process. The mean squared, mean squared log and mean absolute percentage error metrics were considered. 	
The mean absolute percentage error ($\varepsilon$) resulted in consistently lower lossess.
To discuss the performance of the models, the $\varepsilon$ and the corresponding standard deviation ($\sigma$), is presented.
The $\sigma$ quantifies the noise of the models in predicting superficial velocities.
The MFM noise is an important consideration during practical use as gas-liquid flows have greater temporal variations in the flow-field than single phase flows.
The performance of all models is visualized in \cref{fig:results}.
The key takeaways are:
\begin{enumerate}
\item In general, $\varepsilon$ and $\sigma$ is lower in the less complex bubbly flow regime.
\item In general, the performance of \htxt{Model A}, \htxt{B}, \htxt{C} is worse than models with convolutional heads, \htxt{Model D}, \htxt{E}, \htxt{F}.
\item Using a Time Distributed Dense layer, \ie \htxt{~Model B}, provides better performance than larger models (\ie \htxt{~Model A}) or models with a recurrent layer (\ie \htxt{ Model C}).
\item {\helvet Model F} almost systematically underperforms its much leaner counterpart, {\helvet Model E}. The implication is that having separate kernels for each superficial velocity does not result in an improvement.
\item The $\varepsilon$ at extreme values of $\alpha$ is remarkably low $\left(<7.5\%\right)$ for {\helvet Model E}. Omitting an LSTM tail (\ie \htxt{~Model D}) results in a significant performance penalty.
\item Overall, the best performer is {\helvet Model E} whose $\varepsilon$ remains below 5\% for the bubbly flow regime and reaches up to 7.5\% for slug and churn-turbulent flow regimes; the $\sigma$ remains below 10\% for a majority of the flow conditions.
\end{enumerate}

The results have significant implications towards the application of neural networks in the inference of gas-liquid flowrates. 
The study demonstrates that a \textit{single} neural network can be trained to infer flowrates for \textit{multiple} flow regimes; it is unnecessary to first compartmentalize flow by regimes or other statistical parameters, and then require a separate neural network for each.
The study shows that, apart from basic scaling of the input data, there is no additional pre-processing needed.
Lastly, the improvement in performance when a convolutional head is utilized is significant (traditional ANNs have $\varepsilon>15\%$ at high $\alpha$).
This may allude to the reason why early studies may have had less success in directly using ANNs or relied on additional pre-processing methods.

\subsection{Convolutional Feature Maps}

\begin{figure}[!ht]
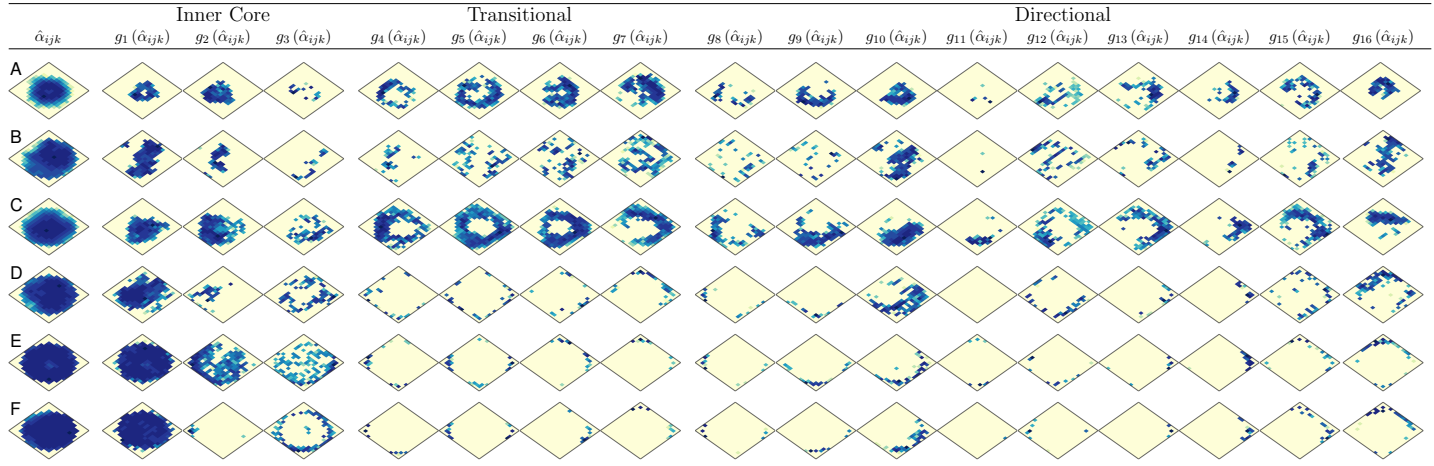

\setlength{\tabcolsep}{0pt}
\resizebox{\linewidth}{!}{
\begin{tabular}{cp{10pt}cccp{10pt}ccccp{10pt}ccccccccc}
\hline
& & \multicolumn{3}{c}{\Large Inner Core} & & \multicolumn{4}{c}{\Large Transitional} & & \multicolumn{9}{c}{\Large Directional} \T\\
$\hat{\alpha}_{ijk}$ & & \gk{1} & \gk{2} & \gk{3} & & \gk{4} & \gk{5} & \gk{6} & \gk{7} & & \gk{8} & \gk{9} & \gk{10} & \gk{11}& \gk{12} & \gk{13} & \gk{14} & \gk{15} & \gk{16} \B\\
\hline \\[-0.5em]
\voidplots{B}{A} \T\\
\voidplots{A}{B} \\
\voidplots{D}{C} \\
\voidplots{C}{D} \\
\voidplots{F}{E} \\
\voidplots{E}{F} \\
\hline
\end{tabular}
}
\caption{Feature maps of the first convolutional layer of \htxt{Model E}. Each 2D map is a result of averaging in the $i$ direction. The first column presents the input experimental matrix, and proceeding columns present kernel output. Rows represent a specific $(v_g, v_f)$ where the same alphabetical symbols in \cref{fig_expdata} are used. Scale of each map is logarithmic and self-normalized.}\label{fig:conv3dmap}
\end{figure}

Why are architectures with a leading convolutional layer better at inference of superficial velocities?
Convolutional layers have several properties (sparse interactions, parameter sharing, and translational invariance \cite[\ch 9]{goodfellow2016deep}) that result in an efficient method of detecting patterns.
For \htxt{Model E}, the first Conv3D layer has 16 kernels, $g_n$, each with a size of $(18,3,3)$. 
The feature maps, $g_n\left(\hat{\alpha}_{ijk}\right)$, for varying flow regimes are presented in \cref{fig:conv3dmap}.
The maps visualize how the network decomposes $\hat{\alpha}_{ijk}$ for downstream layers.
The maps are categorized into: Inner Core, Transitional and Directional.
The Inner Core kernels ($g_{1\rightarrow3}$), extract internal features of the flow.
The Transitional kernels ($g_{4\rightarrow7}$) retain aggregated sectors of the flow in approximately four quadrants.
The Directional kernels ($g_{8\rightarrow16}$) provide smaller isolated features.
It is interesting to see that for regimes that completely dominate the pipe cross-section (such as Experiment \htxt{E} and \htxt{F}), most of the regression information comes from the Inner Core kernels.
Whereas less dominating flows (\eg Experiment \htxt{B} and \htxt{C}), forward information from almost all kernels.

Physically, flow regimes have dominating features (\eg slug flow with periodic Taylor bubbles) that have allowed manual classification and formation of regime maps in the past \cite[\ch 4]{yadigaroglu2017introduction}.
The results from this study suggest that within each flow regime, there are imperceptible spatiotemporal patterns that aid in inferring precisely where in $(v_g,v_f)$ space the observed flow is.
The outcome demonstrates the supremacy of neural networks in such non-linear regression problems. 
Supplementary videos showcasing the sequential activation of all layers are \href{https://github.com/a-jd/bnn/tree/master/videos}{available}.

\subsection{Practical Considerations}

The results indicate that the WMS, coupled directly to a neural network, can provide accurate and precise inference of superficial velocities in gas-liquid flows. 
The implementation of the system in practice requires a few considerations.
The first consideration is whether the neural network side would be able to process the WMS void fraction frames quickly enough to ensure a responsive meter.
The average time taken to process a 512-frame (or \SI{204.8}{\milli\second}) WMS measurement is presented in \cref{table:gputimes}.
For \htxt{Model E}, this value is \SI{3.36}{\milli\second}, roughly two orders of magnitude lower than the time-frame analyzed. 
Therefore, if the time taken for the WMS data to be acquired and transmitted to the neural network is $<\SI{100}{\milli\second}$, the overall meter response frequency would be on the order of $\SI{10}{\hertz}$. 
As the neural network execution time is, apart from the model architecture, hardware and software dependent, any future improvement in either would result in an improvement in response time.

\begin{table}[!ht]
\caption{Average time taken for a $(512,16,16)$ $\hat{\alpha}_{ijk}$ matrix to be evaluated.}
\label{table:gputimes}
\begin{tabular}{cc}
\hline
& Time per $\hat{\alpha}_{ijk}$ $\left[\SI{}{\milli\second}\right]$ \\
\hline
{\helvet Model A} & 0.45 \\
{\helvet Model B} & 0.47 \\
{\helvet Model C} & 18.83 \\
{\helvet Model D} & 0.94 \\
{\helvet Model E} & 3.36 \\
{\helvet Model F} & 31.50 \\
\hline
\end{tabular}
\end{table}

Apart from the hardware required for the WMS \cite{prasser1998new}, additional hardware requirements to add the neural network flow meter capability are straightforward. 
A computer with the capability to run \pkg{Tensorflow} is needed. 
This study used a desktop with a single \textit{Intel} 9900K CPU and a single \textit{NVIDIA} TITAN RTX GPU. 
Less powerful hardware may be used if the response time can be relaxed.
Lastly, if any changes in the loop geometry, WMS placement location ($L/D$), WMS specifications, and loop operating conditions occur -- the neural network would require a `re-calibration', in the sense that new training data would be needed.

\section{Concluding Remarks}

The study finds that neural networks provide an accurate and precise method to infer the superficial velocities for gas-liquid flows in a small-diameter vertical pipe using wire-mesh sensors (WMS).
The sample database has a superficial velocity range scaling approximately two orders of magnitude for both liquid and gas, spanning multiple flow regimes.
Apart from linear transformations and sampling of the raw experimental void fraction matrices, no additional pre-processing methods (\eg statistical filtering, binning, flow regime classification, \etc) were required. 
The major findings are:
\begin{enumerate}
\item The study demonstrates that a \textit{single} neural network model can be used to span multiple gas-liquid flow regimes, while maintaining good accuracy and precision. Multiple models for each flow regime is not necessary.
\item Architectures that implement combinations of Conv3D and LSTM layers outperform pure ANNs, CNNs, or RNNs. 
The finding indicates that the architecture proposed provides a significant advantage in regressing the complex 3D features exhibited in gas-liquid flows and their temporal evolution. 
For \htxt{Model E} (\cref{fig:arch}), the mean absolute percentage error ($\varepsilon$) remains below 5\% for the bubbly flow regime and reaches up to 7.5\% for slug and churn-turbulent flow regimes. 
The noise in inference of superficial velocities (quantified by the standard deviation of $\varepsilon$) remains below 10\% for a majority of the flow conditions.
\item The deployment of the proposed two-phase flow meter is practical. 
The amount of time required to evaluate a void fraction matrix is roughly two orders of magnitude lower than the corresponding matrix's temporal length.
\item The study demonstrates that the addition of the LSTM layer provides a 5-10\% decrease in $\varepsilon$ against fully-connected layers (\eg \htxt{~Model D} \vs \htxt{~Model E}). 
This outcome is prominent in the slug flow regime where features have a strong temporal pattern.
\item The methodology proposed produces a two-phase mass flow meter that is about two orders of magnitude more responsive than current methods (\cref{table_summ} where the sampling time-period, $T$, is tabulated for each approach).
\end{enumerate}

There are several follow-up tasks:
\begin{enumerate}
\item A real-time demonstration of the two-phase flow meter is needed. This would be straightforward as the hardware and software for driving the WMS and neural network are readily available and deployable. The only unknown is the transfer time for the WMS data to be sent to the neural network.
\item The current data set is constrained to 46 permutations of $(v_g,vf)$ that are \SI{10}{\second} (or 25,000 frames) each. Increasing the training dataset volume may lead to a more accurate or more precise model. Increasing the dataset operating envelope into annular flow would also be interesting. This task would require a significant experimental campaign.
\item The current dataset only uses data at a single flow loop location ($L/D=151$). Having multiple WMS at varying $L/D$ would allow investigations of whether the development of gas-liquid flows as a concurrent input would improve model performance. 
\item Although several hyperparameters were probed in a systematic manner, an exhaustive exploration of the entire parameter-space was not undertaken. A tool such as \pkg{HyperOpt} \cite{Bergstra2013} could be used to further refine the models.
\end{enumerate}

\section*{Acknowledgments}
The TOPFLOW experiments and WMS post-processing algorithms were carried out within the framework of research projects funded by the German Federal Ministry for Economic Affairs and Energy, project numbers: 1501411 and 1501329.


\bibliographystyle{ieeetr}
\bibliography{mfnn} 

\begin{thebibliography}{10}

\bibitem{Thorn2013}
R.~Thorn, G.~A. Johansen, and B.~T. Hjertaker, ``{Three-phase flow measurement
  in the petroleum industry},'' {\em Measurement Science and Technology},
  vol.~24, no.~1, 2013.

\bibitem{Ricard2005}
F.~Ricard, C.~Brechtelsbauer, X.~Y. Xu, and C.~J. Lawrence, ``{Monitoring of
  multiphase pharmaceutical processes using electrical resistance
  tomography},'' {\em Chemical Engineering Research and Design}, vol.~83, no.~7
  A, pp.~794--805, 2005.

\bibitem{viswanathan1989damage}
R.~Viswanathan, {\em Damage mechanisms and life assessment of high temperature
  components}.
\newblock ASM international, 1989.

\bibitem{todreas2011nuclear}
N.~E. Todreas and M.~S. Kazimi, {\em Nuclear Systems Volume I: Thermal
  Hydraulic Fundamentals}.
\newblock CRC press, 2011.

\bibitem{Oddie2004}
G.~Oddie and J.~A. Pearson, ``{Flow-rate Measurement in Two-phase Flow},'' {\em
  Annual Review of Fluid Mechanics}, vol.~36, no.~1, pp.~149--172, 2004.

\bibitem{Rajan1993}
V.~S.~V. Rajan, R.~K. Ridley, and K.~G. Rafa, ``{Multiphase Flow Measurement
  Techniques -- A Review},'' {\em Journal of Energy Resources Technology},
  vol.~115, 1993.

\bibitem{Chaouki1997}
J.~Chaouki, F.~Larachi, and M.~P. Dudukovi{\'{c}}, ``{Noninvasive Tomographic
  and Velocimetric Monitoring of Multiphase Flows},'' {\em Industrial {\&}
  Engineering Chemistry Research}, vol.~36, pp.~4476--4503, nov 1997.

\bibitem{Heindel2011}
T.~J. Heindel, ``{A review of X-ray flow visualization with applications to
  multiphase flows},'' {\em Journal of Fluids Engineering, Transactions of the
  ASME}, vol.~133, no.~7, 2011.

\bibitem{Ismail2005}
I.~Ismail, J.~C. Gamio, S.~F.~A. Bukhari, and W.~Q. Yang, ``{Tomography for
  multi-phase flow measurement in the oil industry},'' {\em Flow Measurement
  and Instrumentation}, vol.~16, pp.~145--155, 2005.

\bibitem{Cai1994}
S.~Cai, H.~Toral, J.~Qiu, and J.~S. Archer, ``{Neural network based objective
  flow regime identification in air-water two phase flow},'' {\em The Canadian
  Journal of Chemical Engineering}, vol.~72, pp.~440--445, jun 1994.

\bibitem{Tsoukalas1997}
L.~H. Tsoukalas, M.~Ishii, and Y.~Mi, ``{A neurofuzzy methodology for
  impedance-based multiphase flow identification},'' {\em Engineering
  Applications of Artificial Intelligence}, vol.~10, pp.~545--555, dec 1997.

\bibitem{Mi1998}
Y.~Mi, M.~Ishii, and L.~Tsoukalas, ``{Vertical two-phase flow identification
  using advanced instrumentation and neural networks},'' {\em Nuclear
  Engineering and Design}, vol.~184, pp.~409--420, aug 1998.

\bibitem{Mi2001}
Y.~Mi, M.~Ishii, and L.~Tsoukalas, ``{Flow regime identification methodology
  with neural networks and two-phase flow models},'' {\em Nuclear Engineering
  and Design}, vol.~204, pp.~87--100, feb 2001.

\bibitem{Wu2001}
H.~Wu, F.~Zhou, and Y.~Wu, ``{Intelligent identification system of flow regime
  of oil–gas–water multiphase flow},'' {\em International Journal of
  Multiphase Flow}, vol.~27, pp.~459--475, mar 2001.

\bibitem{Hernandez2006}
L.~Hern{\'{a}}ndez, J.~E. Juli{\'{a}}, S.~Chiva, S.~Paranjape, and M.~Ishii,
  ``{Fast classification of two-phase flow regimes based on conductivity
  signals and artificial neural networks},'' {\em Measurement Science and
  Technology}, vol.~17, no.~6, pp.~1511--1521, 2006.

\bibitem{Julia2008}
J.~E. Juli{\'{a}}, Y.~Liu, S.~Paranjape, and M.~Ishii, ``{Upward vertical
  two-phase flow local flow regime identification using neural network
  techniques},'' {\em Nuclear Engineering and Design}, vol.~238, pp.~156--169,
  jan 2008.

\bibitem{Wang2017}
L.~Wang, J.~Liu, Y.~Yan, X.~Wang, and T.~Wang, ``{Gas-Liquid Two-Phase Flow
  Measurement Using Coriolis Flowmeters Incorporating Artificial Neural
  Network, Support Vector Machine, and Genetic Programming Algorithms},'' {\em
  IEEE Transactions on Instrumentation and Measurement}, vol.~66, pp.~852--868,
  may 2017.

\bibitem{Meribout2010}
M.~Meribout, N.~Al-Rawahi, A.~Al-Naamany, A.~Al-Bimani, K.~Al-Busaidi, and
  A.~Meribout, ``{Integration of impedance measurements with acoustic
  measurements for accurate two phase flow metering in case of high
  water-cut},'' {\em Flow Measurement and Instrumentation}, vol.~21, no.~1,
  pp.~8--19, 2010.

\bibitem{Shaban2014}
H.~Shaban and S.~Tavoularis, ``{Measurement of gas and liquid flow rates in
  two-phase pipe flows by the application of machine learning techniques to
  differential pressure signals},'' {\em International Journal of Multiphase
  Flow}, vol.~67, pp.~106--117, dec 2014.

\bibitem{Shaban2015}
H.~Shaban and S.~Tavoularis, ``{The wire-mesh sensor as a two-phase flow
  meter},'' {\em Measurement Science and Technology}, vol.~26, p.~015306, jan
  2015.

\bibitem{Fan2014}
S.~Fan and T.~Yan, ``{Two-phase air-water slug flow measurement in horizontal
  pipe using conductance probes and neural network},'' {\em IEEE Transactions
  on Instrumentation and Measurement}, vol.~63, pp.~456--466, feb 2014.

\bibitem{prasser1998new}
H.-M. Prasser, A.~B{\"o}ttger, and J.~Zschau, ``A new electrode-mesh tomograph
  for gas--liquid flows,'' {\em Flow measurement and instrumentation}, vol.~9,
  no.~2, pp.~111--119, 1998.

\bibitem{prasser2007construction}
H.~Prasser, D.~Lucas, M.~Beyer, C.~Vallee, E.~Krepper, T.~Hohne, A.~Manera,
  H.~Carl, H.~Pietruske, P.~Schutz, {\em et~al.}, ``Construction and execution
  of experiments at the multi-purpose thermal hydraulic test facility topflow
  for generic investigations of two-phase flows and the development and
  validation of cfd codes,'' {\em Wissenschaftlich-Technische Berichte,
  FZD-481}, 2007.

\bibitem{Kaichiro1984}
M.~Kaichiro and M.~Ishii, ``{Flow regime transition criteria for upward
  two-phase flow in vertical tubes},'' {\em International Journal of Heat and
  Mass Transfer}, vol.~27, no.~5, pp.~723--737, 1984.

\bibitem{prasser2005influence}
H.-M. Prasser, M.~Beyer, A.~B{\"o}ttger, H.~Carl, D.~Lucas, A.~Schaffrath,
  P.~Sch{\"u}tz, F.-P. Weiss, and J.~Zschau, ``Influence of the pipe diameter
  on the structure of the gas-liquid interface in a vertical two-phase pipe
  flow,'' {\em Nuclear Technology}, vol.~152, no.~1, pp.~3--22, 2005.

\bibitem{goodfellow2016deep}
I.~Goodfellow, Y.~Bengio, and A.~Courville, {\em Deep learning}.
\newblock MIT press, 2016.

\bibitem{chollet2015keras}
F.~Chollet {\em et~al.}, ``Keras.'' \url{https://keras.io}, 2015.

\bibitem{tensorflow2015}
M.~Abadi {\em et~al.}, ``{TensorFlow}: Large-scale machine learning on
  heterogeneous systems,'' 2015.
\newblock Software available from tensorflow.org.

\bibitem{Hahnioser2000}
R.~H. Hahnioser, R.~Sarpeshkar, M.~A. Mahowald, R.~J. Douglas, and H.~S. Seung,
  ``{Digital selection and analogue amplification coexist in a cortex- inspired
  silicon circuit},'' {\em Nature}, vol.~405, pp.~947--951, jun 2000.

\bibitem{yadigaroglu2017introduction}
G.~Yadigaroglu and G.~F. Hewitt, {\em Introduction to multiphase flow: basic
  concepts, applications and modelling}.
\newblock Springer, 2017.

\bibitem{Bergstra2013}
J.~Bergstra, D.~Yamins, and D.~D. Cox, ``{Making a science of model search:
  Hyperparameter optimization in hundreds of dimensions for vision
  architectures},'' in {\em 30th International Conference on Machine Learning,
  ICML 2013}, vol.~28, pp.~115--123, 2013.

\end{thebibliography}
\end{document}